\documentclass[conference,letterpaper]{IEEEtran}

\RequirePackage{acronym,amssymb,amsthm,dsfont,bbding,fancyhdr,graphicx,lastpage}
\usepackage[ruled,vlined]{algorithm2e}
\usepackage{tabularx,diagbox,mathrsfs}
\usepackage[utf8]{inputenc} 
\usepackage{pgf}
\usepackage{stmaryrd}
\usepackage[cmex10]{amsmath} 
\acrodef{ACDIS}[ACDIS]{Adaptive Communication Decision and Information Systems}
\acrodef{AWGN}{additive white gaussian noise}
\acrodef{BER}{bit-error-rate}
\acrodef{BEC}{binary erasure channel}
\acrodef{BPSK}{binary phase-shift keying}
\acrodef{BSC}{binary symmetric channel}
\acrodef{FER}[FER]{frame error rate}
\acrodef{iid}[i.i.d.]{independent and identically distributed}
\acrodef{IoT}[IoT]{Internet of Things}
\acrodef{LDPC}[LDPC]{low-density parity-check}
\acrodef{LLR}[LLR]{log-likelihood ratios}
\acrodef{MLC}[MLC]{multi-level coding}
\acrodef{MLE}[MLE]{maximum eikelihood estimate}
\acrodef{PAC}[PAC]{polarization-Assisted Convolutional}
\acrodef{PM}[PM]{path metric}
\acrodef{BM}[BM]{branch metric}
\acrodef{PMF}[PMF]{probability mass function}
\acrodef{RV}{random variable}
\acrodef{SNR}{signal-to-noise ratio}
\acrodef{SC}{successive cancellation}
\acrodef{SCL}{successive cancellation list}

\newcommand{\calL}{\mathcal{L}}

\newcommand{\calO}{\mathcal{O}}

\newcommand{\intseq}[2]{[{#1};{#2}]}
\newcommand{\eqdef}{\triangleq}

\renewcommand{\geq}{\geqslant} 

\begin{document}
\title{Quantum Precoded Polar Codes} 


\author{%
  \IEEEauthorblockN{Tyler Kann}
  \IEEEauthorblockA{Department of Electrical Engineering \\
                    Georgia Institute of Technology\\
                    Atlanta, Georgia\\
                    kann@gatech.edu}
  \and
  \IEEEauthorblockN{Shrinivas Kudekar}
  \IEEEauthorblockA{kudekar@gmail.com}
  \and
    \IEEEauthorblockN{Matthieu R. Bloch}
  \IEEEauthorblockA{Department of Electrical Engineering \\
                    Georgia Institute of Technology\\
                    Atlanta, Georgia\\
                    matthieu.bloch@ece.gatech.edu}
  \and
}

\maketitle

\begin{abstract}
  We introduce a new family of CSS codes obtained from rate-1 precoded polar codes, which harnesses the precoding benefits obtained for classical short blocklength polar codes. We optimize the rate profile and precoder of these codes with a genetic algorithm, and present codes of dimension $[\![256,2 ]\!]$  and $[\![512,2 ]\!]$ that have logical error rates similar to the $[\![1201,1,25]\!]$ surface code over the depolarizing channel.  
\end{abstract}

\section{Introduction}
Quantum computers are being developed with an increasing number of qubits ($N\geq256)$. However, qubits are an incredibly fragile resource, and quantum computers are still limited by the error on these qubits. Developing quantum error correction codes that mitigate quantum errors is therefore crucial. 

There are many classes of quantum error correcting codes. Two of specific interest for the present work are stabilizer codes \cite{gottesman1997stabilizer}, and CSS codes \cite{calderbankGoodQuantumErrorCorrecting1996a, steaneMultipleparticleInterferenceQuantum1996}, a special type of stabilizer code. Quantum polar codes constructed as CSS codes have previously been explored \cite{renesEfficientPolarCoding2012, goswamiFaulttolerantPreparationQuantum2023, hidakaInterpolationQuantumPolar2025}, most recently under \ac{SCL} decoding \cite{gongImprovedLogicalError2024}. The performance of \cite{gongImprovedLogicalError2024} is competitive with the surface code \cite{bravyiEfficientAlgorithmsMaximum2014}, however the polar codes were constructed primarily to satisfy the CSS constraints; they are not necessarily the best construction. 

Despite classical asymptotic optimality, \mbox{polar codes \cite{Arikan2009}} suffer from shortcomings in the short blocklength regime, when bits have not sufficiently polarized. In recent years, there has been increased interest in developing algorithmic and engineering approaches to boost the performance of polar codes. Of these, most prominent is list decoding \cite{Tal_2015}. Applying a list decoder to the \ac{SC} decoding of polar codes enables the decoder to track multiple paths at once, allowing them to retroactively correct a previous bit decision that would have been myopically made in the absence of a list. Going from \ac{SC} to \ac{SCL} happens at the decoder, meaning no modification needs to be made to the polar code (or the Clifford gates constructing the code). 

Additionally, it has been shown that applying an outer code to a polar code also significantly improves the performance \cite{Fazeli_2019}, even when the outercode is \emph{rate-1} \cite{Li2019, minweight}. The \mbox{rate-1} precoder applied is not acting as an effective outer code (since it is \mbox{rate-1}) but is actually aiding the list decoder, making incorrect decisions easier to detect. This means that a rate-1 precoded polar code (which we will simply refer to as a precoded polar code) is only useful when the decoder has memory (e.g., \ac{SCL}). Precoded polar codes have also been studied under the name polarization-adjusted convolutional (PAC) codes \cite{ArikanPAC} and polar codes with dynamically frozen bits \cite{Trifonov2013}. 

In light of the usefulness of a list decoder for quantum polar code \cite{gongImprovedLogicalError2024}, a list decoder for quantum precoded polar codes is a natural extension. However, the extension is not trivial, as the constraints required to construct a CSS code mean that one cannot use a time-invariant convolutional precoding. Additionally, as seen in the classical case, the information sets constructed to best make use of the precoder are also driven algorithmically \cite{kannPathMetricBased2024, WeightedSum, moradiPACCodeRateProfile2024a}, as opposed to the asymptotically optimal constructions based on information-theoretic quantities. Another challenge to address is therefore ensuring that the sets that best use the precoder also lead to valid CSS constructions.

The contributions of this work are addressing following two challenges:
\begin{enumerate}
    \item Understanding how a rate-1 precoder can be applied to a quantum polar code, and conducting a search for a good precoder.
    \item Conducting a search for an improved information set, that is both valid and makes use of the precoding. 
\end{enumerate}

The rest of the paper is structured as follows. In Section \ref{background}, we provide background review of necessary concepts. In Section \ref{QPPC}, we present a way to construct quantum precoders. Section \ref{eval}, we showcase some of the new codes found in our method and provide additional discussions. We then conclude in Section \ref{conclusion}.

\section{Background}
\label{background}
\subsection{Polar Codes}

A polar code for channel coding is characterized by its blocklength $N = 2^n, n \in \mathbb{N}$, the number of information bits $K$, and a set $\mathcal{A}\subset\intseq{1}{N}$ that specifies where to place information bits. We also define $\mathcal{F} = \mathcal{A}^c$. Specifically, a vector of $K$ information bits $m$  is encoded into a length $N$ vector $u$ such that $u_\mathcal{A} = m$ and $u_{\mathcal{F}} = 0$, the all-zero vector. The sets $\mathcal{A}$ and $\mathcal{F}$ are called the information and frozen set, respectively. For algebraic simplicity, we additionally define $A \in \mathbb{F}_2^{K \times N}$ and $F \in \mathbb{F}_2^{N \times (N-K)}$ as the selection matrices of the sets $\mathcal{A}$ and $\mathcal{F}$, e.g. $mA = u$ and $uF = 0$. Upon setting $G_2 = \left[\begin{smallmatrix} 1 & 0 \\ 1 & 1 \end{smallmatrix}\right]$, the base polarization matrix, and $G = G_2^{\otimes n}$ as the $n$th order Kronecker product of $G_2$, a codeword $x$ is created through the operation $x \eqdef uG$, a process known as the polar transform. The codeword $x$ is then sent over the channel to a receiver that observes a noisy version $y$, e.g., $y_i= x_i+ n_i, n_i \sim \textrm{Bern}(p)$ if the channel is an \acf{BSC} channel with crossover probability $p$. 
  
The standard decoding algorithm for polar codes is the \ac{SC} decoder, by which bits from the information set $\mathcal{A}$ are successively decoded based on their \ac{LLR}, $\lambda$, given the past decoded bits according to the maximum-likelihood rule. 
\begin{align}
\forall i\in\mathcal{A}\quad  \hat{u}_i &=     \begin{cases}
                    0 & \text{if } \lambda_i = \ln \frac{P(y,\hat{u}_{0:i-1}|\hat{u}_i = 0) }{P(y,\hat{u}_{0:i-1}|\hat{u}_i = 1) } > 0\\
                    1 & \text{else.} 
                                         \end{cases}, \label{info} \\
  \forall i\in\mathcal{F}\quad \hat{u}_i&=0, \label{frozen}
\end{align}
where $\hat{u}_{0:i-1} = \{\hat{u}_0, \hat{u}_1, \dots, \hat{u}_{i-1} \}$ denotes the vector of past decisions. The \acp{LLR} can be efficiently computed recursively, resulting in a decoding complexity $O(N\log N)$. The \acp{LLR} can also be viewed as decisions made at the output of individual \emph{bit channels}, corresponding to channels with input bit $u_i$ and output $(y,u_{0:i-1})$. The choice of the set $\mathcal{A}$ plays a crucial role in determining the performance of polar codes. 

\subsection{Successive Cancellation List Decoding}
\label{sec:SCL}

Given that each decision in the \ac{SC} decoder relies on previously decoded bits, one bad decision can propagate and make future bits erroneous. \ac{SCL} decoding~\cite{Tal_2015} attempts to avoid this problem and can be implemented relatively efficiently. In \ac{SCL}, the decoding process is viewed as following branches of a tree, simultaneously tracking up to $L$ branches. The tree splits at every non-frozen bit $i$, creating two branches: $\hat{u}_i = 0$ and $\hat{u}_i = 1$. For every path in the set $\calL$ of currently tracked paths, the \ac{PM} is computed as the accumulation of \acp{BM}. Each \ac{BM} is the penalty based on the decision for both frozen and information bits. The updates are tracked according to: 
\begin{align}
&\forall l\in\calL, \quad  \textrm{PM}_i[l] = \textrm{PM}_{i-1}[l] + \textrm{BM}_i[l] \label{PM}\\ 
& \textrm{BM}_i[l] = f(\hat{u}_i, \lambda_i), \label{BM}
\end{align} 
Where $f(\hat{u}_i, \lambda_i)$ is a penalty function based on the decision $\hat{u}_i$ and the magnitude of certainty $\lambda_i$, e.g., \mbox{$f(\hat{u}_i, \lambda_i) = \log_2(1+2^{-\lambda_i \cdot (-1)^{\hat{u_i}}})$}. 
When $|\calL|$ is greater than $L$, the list is pruned back to only $L$ paths, keeping those with the lowest path metrics. 

\subsection{Syndrome Decoding of Polar Codes}
\label{syndrome}
A key challenge for quantum systems is that the data cannot be measured without disturbing the state. Therefore, when receiving a received noisy codeword $y$, instead of trying to estimate $x$, the decoder may only access information on the noise, $n$, through syndrome measurements. The decoder now outputs an estimate of the noise, $\hat{n}$, and reapplies it as $y + \hat{n} = x + n + \hat{n} = \hat{x}$, ideally removing the noise. This setup is known as syndrome decoding, which can be viewed as source coding without side information \cite{cronieLosslessSourceCoding2010}. For any code $\mathcal{C}$ that has a parity check matrix $H$, it is known that $xH^T = 0 \forall x \in \mathcal{C}$. Therefore $yH^T = 0 + nH^T = s$, where $s$ is the syndrome of the noise and aids the decoder in estimating $\hat{n}$. For a polar code with information set $\mathcal{A}$, since $G$ is its own inverse, we know that $xGF = uF = 0$, making the parity check matrix $F^TG^T$. The syndrome is obtained as $nGF = s_\mathcal{F}$. The decoder is now aiming to build the entire $N$ length vector $s$, and ultimately obtain an estimate $\hat{n} = sG$. Algorithmically, this means that the decoder is no longer forcing a frozen bit to be $0$ in \eqref{frozen}, but rather the appropriate bit from $s_\mathcal{F}$. In this problem, the decoder sets $y_i = 0 \phantom{.} \forall i $,  a choice that communicates that the decoder has no side information, and preserves the BSC($p$) noise structure of the channel \cite{cronieLosslessSourceCoding2010}.

\subsection{Precoded Polar Codes}

\label{sec:pac-codes}

A precoded polar code still polarizes $u$ to get the final codeword, except now $u = vT$, with $v$ being the new information vector and $T$ being a causal, upper-triangular matrix. The information set $\mathcal{A}$ now affects $v$, having $v_{\mathcal{A}} = m, v_{\mathcal{F}} = 0$. Decoding and metric decisions from \mbox{\eqref{info}-\eqref{BM}} are still based on $\hat{u}_i$, but the decoder's branches are based on $\hat{v}_i$ and $\hat{u}_i$ is now determined from $\hat{v}_{0:i}$ and $T$.  Row $i$ of $T$ determines how the decision of $\hat{v}_i$ affects future decisions of $\hat{u}_{i+1:N-1}$, and column $j$ of $T$ determines how previous bits $\hat{v}_{0:j-1}$ affect $\hat{u}_j$ based on the decision of $\hat{v}_j$. Further discussions of precoded polar codes can be found in \cite{kannPathMetricBased2024, RowshanSCL}.

\subsection{Stabilizer Codes and CSS Codes}

A stabilizer code, \mbox{$H = [H_X | H_Z] \in \mathbb{F}_2^{(N-K) \times 2N}$}, must satisfy
\begin{align}
    \langle H,H\rangle = H_XH_Z^T + H_ZH_X^T = 0 \textrm{ mod }2, \label{symp}
\end{align}
where $\langle \cdot,\cdot \rangle$ denotes the symplectic inner product. A specialized stabilizer code is the CSS code, constructed from two classical codes $\mathcal{C}_X$ and $\mathcal{C}_Z$ and their parity check matrices $\tilde{H}_X$ and $\tilde{H}_X$. We now define $H_X = \begin{bmatrix} 0 \\ \tilde{H}_X\end{bmatrix}$ and $H_Z = \begin{bmatrix} \tilde{H}_Z \\ 0 \end{bmatrix}$. This reduces \eqref{symp} to $\tilde{H}_X\tilde{H}_Z^T = 0$, a requirement that can also expressed as $\mathcal{C}_Z^\perp \subseteq \mathcal{C}_X$, where $\mathcal{C}_Z^\perp$ is the dual of $\mathcal{C}_Z$. If $\mathcal{C}_X$ and $\mathcal{C}_Z$ are of dimension $[N,K_X]$ and $[N,K_Z]$, respectively, the resulting CSS code is a quantum code of dimension \mbox{$\llbracket N, K_X + K_Z -N\rrbracket$}. 

We limit the scope of the present paper to the depolarizing channel. The depolarizing channel, characterized by parameter $p\in[0,1]$ is a channel that can affect each qubit by a Pauli operator $X,Z,Y$, each with probability $\frac p3$. Using the same $[X|Z]$ notation as above, the Pauli operators $X,Z,Y$ can be mapped to binary, with $Y$ occurring when both an $X$ and a $Z$ error occur. In some cases, $n$ will refer to Pauli noise, $n = [n_X|n_Z] \in \mathbb{F}_2^{2N}$, however in some cases $n$ still references a binary, classical noise, a distinction that will hopefully be unambiguous given the context. $X$ errors in $n_X$ are measured by $H_Z$ and $Z$ errors in $n_Z$ are measured by $H_X$. 

\subsection{Quantum Polar Codes}
\label{quantumpc}
The quantum encoding circuit for $\mathcal{C}_Z$  is identical to a classical polar code, replacing the classical CNOT gate with a quantum CNOT gate. Upon doing a Hadamard transform, the CNOT gates are reversed (meaning the roles of control and target are reversed), and thus the polarization for $\mathcal{C}_X$ is performed in the reverse direction. This means that while $\mathcal{C}_Z$ is a traditional polar code with generator matrix $G$ (and sets $\mathcal{A}_Z, \mathcal{F}_Z$), $\mathcal{C}_X$ is a reversed polar code with generator matrix $G^T$ (and sets $\mathcal{A}_X, \mathcal{F}_X)$. With parity check matrices $\tilde{H}_Z = F_Z^TG^T$ and $\tilde{H}_X = F_X^TG$, \eqref{symp} reduces to 
\begin{align}
    F_X^TGGF_Z = F_X^TF_Z = 0, 
\end{align}
which can be satisfied when $F_X \cap F_Z = \emptyset$, meaning no bits are simultaneously frozen in both codes. Similar to \cite{gongImprovedLogicalError2024}, both polar codes have $K = \frac N2 + 1$, meaning the resulting quantum polar code is of dimension $\llbracket N,2\rrbracket$. The logical information is contained in $\mathcal{A}_X \cap \mathcal{A}_Z$, indices that have information encoded in both bases. We call this intersection the logical bits. The remaining information bits (which we call stabilizer bits) are still decoded as information bits, but a decoding success only requires that the logical bits be decoded correctly. 

The reversed polar code is polarized by the matrix $G^T$, which can be decomposed to $JGJ$, where $J$ is the exchange matrix, 
\begin{align}
J_n &= \begin{bmatrix}
0 & 0& \cdots & 0 & 1 \\
0 & 0 & \cdots & 1 & 0 \\
\vdots & \vdots & \ddots & \vdots & \vdots \\
0 & 1 & \cdots & 0 & 0 \\
1 & 0 & \cdots & 0 & 0
\end{bmatrix}.
\end{align}

This means that $\tilde{H}_X^T = G^TF_X = JGJF_X = JGF_Z$, if the two codes are equivalent up to the reversal. When receiving a noisy quantum codeword, the decoder can use the ancilla qubits to measure the $X$ syndrome $s^X_{\mathcal{F}_Z} = n_XGF_Z$, change bases via the Hadamard transform, and subsequently measure \mbox{$s^Z_{\mathcal{F}_Z} = n_ZJGF_Z$}, the syndrome of the \emph{reversal} of $n_Z$ but with frozen set $\mathcal{F}_Z$. The decoder needs to account for the reversal, but the benefit is that both codes share the same frozen set, and can be viewed as a joint, \emph{quaternary} decoder, as opposed to two binary decoders. While the results of Slepian-Wolf \cite{Slepian_1973} tell us that the asymptotic performance should be the same, a joint decoder significantly outperforms a two-stage decoding approach. most likely due to the multi-stage error propagation from the stabilizer bits.  We slightly abuse notation and drop the $X/Z$ subscript when referencing the joint code, e.g., we write $\mathcal{A}, \mathcal{F},s$. 

The decoder must now keep track of the quadruple $(p_I, p_X, p_Z, p_Y)$ instead of the scalar \ac{LLR} (and update the checknode and repetition node functions accordingly). Additionally, the list decoder will now branch $4$ times, exploring possibilities of $s_i$ being any of the $4$ Paulis. On each branch $b$, the \ac{BM} function from $\eqref{BM}$ is defined as \mbox{$\textrm{BM}_i[l] =\log(p_b), b \in \{I,X,Z,Y\}$}. For more implementation details, we refer the reader to \cite{gongImprovedLogicalError2024}.

\section{Quantum Precoded Polar Codes}
\label{QPPC}

In this section we establish that, under the correct constraints, a polar code can remain CSS when precoded with matrix $T$, and how additional Clifford gates create this precoding. We then propose a genetic algorithm to search for good quantum precoded polar codes. 

\subsection{Valid Quantum Precoders}
\label{valid}

Similar to the translation of a classical polar encoding circuit to a quantum polar encoding circuit, the quantum precoder $T$ can be created by placing CNOT gates before the CNOT gates implenting $G$. The non-zero element of $T_{i,j}$ can be constructed with a CNOT gate between qubits $i$ and $j$. However, after the Hadamard transform, the CNOT gates are reversed, and now represent $T_{j,i}$. Accounting for this means the precoder must be invariant to reversals, constraining $T$ to be \emph{persymmetric}, i.e., $T=JTJ$. Additionally, $T$ must also be an involution, a requirement to recreate the parity check matrices in Section \ref{syndrome}. With this structure, our new parity check matrices are $\tilde{H}_Z = F_Z^TT^TG^T$ and $\tilde{H}_X = F_X^TTG$, and \eqref{symp} becomes: 
\begin{align}
    F_X^TTGGTF_Z = F_X^TF_Z = 0.
\end{align} Moreover, the constraint
\begin{align}
    \tilde{H}_X^T = G^TT^TF_X = JGJJTJF_X = JGTF_Z,
\end{align} means the algorithmic procedure described in Section \ref{quantumpc} holds. Thus, the decoder can employ a joint quaternary decoder for our quantum precoded polar code. Note that, now, \mbox{$s^X = n_X\tilde{H}_Z^T = n_XGTF_Z$}, and polarization occurs \emph{before} any precoding. This has been explored before in \cite{kannSourcePolarizationAdjustedConvolutional2023b}, in which the authors apply classical precoded polar codes to the problem of source coding with side information. As stated, since syndrome decoding is a specialization of source coding, it is appropriate to use this order of operations, allowing for channel coding at the encoder and source coding at the decoder. Since we have no side information, we do not need to account for the permutation.

The decoder also tracks an auxiliary variable $r \eqdef nG$, and branches of $s$ influence $r$, the polarized source. The roles of $s$ and $r$ analogous to those of $v$ and $u$ in Section \ref{sec:pac-codes}. The benefit of the joint decoder is that one index is keeping track of the entire quadruple $(p_I, p_X, p_Z, p_Y)$. However, Pauli decisions of $s$ are still kept as $[s^{X}_i, s^{Z}_i]$, two binary values. This means that the role $T$ plays between \mbox{$s$ and $r$} can be decoupled to functions involving \mbox{$s^X$ and $r^X$} and \mbox{$s^Z$ and $r^Z$}, both behaving independently and identically to the behavior in Section \ref{sec:pac-codes}.

\subsection{Genetic Algorithm Search}

While Section \ref{valid} establishes the functionality and existence of a valid $T$, and Section \ref{quantumpc} establishes that a valid $\mathcal{A}$ exists, finding an optimized $\mathcal{A}$ or $T$ under the CSS constraints is an open problem. Even in the classical regime, the idea of an optimized $T$ is rarely addressed, as typically a time-invariant convolutional matrix is more than sufficient; however these matrices are not their own inverses, making the need for a non-trivial $T$ relevant. Furthermore, \cite{renesAlignmentPolarizedSets2016} showed that, on non-erasure channels, after polarization, a bad bit-channel $i$ does not enforce a good bit-channel $N-1-i$, a necessary condition to have vanishing error under a CSS construction. Therefore, finding high-performance sets using metrics besides mutual information is of interest, e.g. minimum weight \cite{rowshanConvolutionalPrecodingPAC2021}, PM \cite{kannPathMetricBased2024}, or data-driven genetic algorithms \cite{moradiPACCodeRateProfile2024a, moradiMonteCarloBasedConstruction2021}.

A genetic algorithm is a promising approach. A large benefit is that such algorithms are empirically driven, meaning that there are no proxy metrics and a code can be scored base on its actual performance. We propose the following genetic algorithm, detailed in Algorithm \ref{alg:joint} and operating essentially as follows: 
\begin{enumerate}
    \item Use a Genetic Algorithm  ($\mathrm{Set\_GA}$) to create $\mathscr{A}$, a set of high performing information sets, constrained to $F^TJF = 0$.
    \item For each $\mathcal{A} \in \mathscr{A}$, use a Genetic Algorithm  ($\mathrm{T\_GA}$) to optimize $T$, constrained to $TJT^TJ = I$. The set of pairs $(\mathcal{A},T)$ is stored in $\mathscr{P}$.
    \item Alternate Genetic Algorithms to further refine each pair $(\mathcal{A},T) \in \mathscr{P}$, constrained to $F^TJF = 0, TJT^TJ = I$.
\end{enumerate}

A hard problem for any algorithmic construction is simply that the search space is incredibly large. While the constraints imposed by \eqref{symp} prohibit choosing an optimal information set, the search space of \emph{all} valid rate profiles becomes greatly reduced. The indices are now grouped into pairs $(0,N-1) \dots (\frac N2 -1, \frac N2)$. The pairs can either be the logical bits, having both indices in $\mathcal{A}$, or can have exactly one indices in $\mathcal{A}$. We also use a similar approach to \cite{moradiPACCodeRateProfile2024a} where some indices are forced to be frozen, with an additional benefit being that just as many indices are forced to be information bits.

\begin{algorithm}[h]
\DontPrintSemicolon
\SetKwInOut{KwIn}{Input}
\SetKwInOut{KwOut}{Output}
\SetKwFunction{SetGA}{Set\_GA}
\SetKwFunction{TGA}{T\_GA}
\SetKwFunction{RowGA}{Row\_GA}
\SetKw{KwRet}{return}

\KwIn{$N$, $K$, initial frozen set $\mathcal{A}_0$, initial precoder $T_0 = I$ , list size $L$, channel parameter $p$}
\KwIn{Population size $P$, number of offspring $C$, outer iterations $O$}
\KwOut{Optimized pair $(A^\star,T^\star)$}

$\mathscr{A} \gets \SetGA(\mathcal{A}_0, T_0)$\;
$\mathscr{P} \gets \emptyset$\;
\ForEach{$\mathcal{A} \in \mathscr{A}$}{
    $T \gets \TGA(\mathcal{A}, T_0)$\;
    insert $(\mathcal{A},T)$ into $\mathscr{P}$\;
}

\For{$o \gets 1$ \KwTo $O$}{
    Select an elite pair $(\mathcal{A}_{\mathrm{elite}},T_{\mathrm{elite}})$ from $\mathscr{P}$\;
    \tcp{We select from $\mathscr{P}$ and randomly choose if we optimize $T$ or $\mathcal{A}$}
    \uIf{precoder refinement is chosen}{
        $T_{\mathrm{new}} \gets \TGA(\mathcal{A}_{\mathrm{elite}}, T_{\mathrm{elite}})$\;
        insert $(\mathcal{A}_{\mathrm{elite}},T_{\mathrm{new}})$ into $\mathscr{P}$\;
    }
    \Else{
        $\mathscr{A}_{\mathrm{new}} \gets \SetGA(\mathcal{A}_{\mathrm{elite}}, T_{\mathrm{elite}})$\;
        \ForEach{$\mathcal{A}_{\mathrm{new}} \in \mathscr{A}_{\mathrm{new}}$}{
            insert $(\mathcal{A}_{\mathrm{new}},T_{\mathrm{elite}})$ into $\mathscr{P}$\;
            $T_{\mathrm{new}} \gets \RowGA(\mathcal{A}_{\mathrm{elite}}, T_{\mathrm{elite}}, \mathcal{A}_{\mathrm{new}})$\; \tcp{A small version of \TGA focused on $\mathcal{A}_{\textrm{new}} \setminus \mathcal{A}_{\textrm{elite}}$}
            insert $(A_{\mathrm{new}},T_{\mathrm{new}})$ into $\mathscr{P}$\;
        }
    }

    Retain the top $P$ pairs in $\mathscr{P}$ according to their logical error rate sampled from Monte Carlo\;
}
\Return the best pair $(A^\star,T^\star)$ in $\mathscr{P}$\;
\caption{Joint optimization of quantum precoded polar codes}
\label{alg:joint}
\end{algorithm}

\section{Evaluation and Discussion}
\label{eval}

\begin{figure*}[h]
    \centering
    {\input{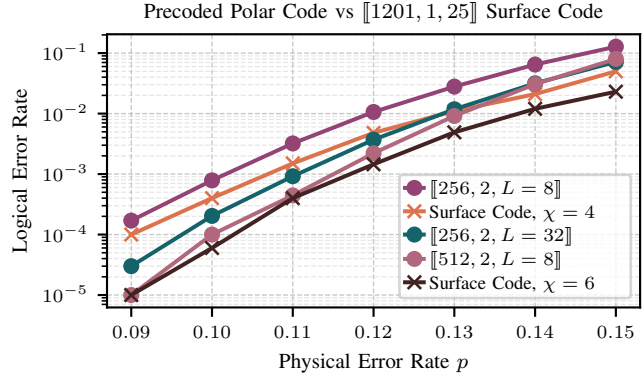}}
    \vspace{-3mm}
    \caption{A comparison between \mbox{$\llbracket N, 2  \rrbracket$} precoded polar codes for $N=256,512$ and $\llbracket 1201, 1, 25\rrbracket$ surface code with $\chi = 4,6$. The \mbox{$\llbracket 256, 2, L  = 8,32 \rrbracket $} codes are two different codes from Algorithm \ref{alg:joint}, as opposed to one code evaluated at two list sizes.}
    \label{surfaceCodeComp}
\end{figure*}

To showcase the performance of this new family of codes, we compare against the $\llbracket1201, 1, 25\rrbracket$ surface code on the depolarizing channel for of $\chi = 4,6$, which correspond to different decoding complexities \cite{bravyiEfficientAlgorithmsMaximum2014}. Remarkably, as seen in Figure \ref{surfaceCodeComp}, we are able to obtain similar raw performance in terms of logical error rate, using only $\llbracket256,2,L=32\rrbracket$ and $\llbracket512,2,L=8\rrbracket$ precoded polar codes, giving nearly a $10$x boost in rate with only one fifth of the qubits needed. Furthermore, an efficient decoder of the surface code can be done with run time $\calO(n \chi^3)$ \cite{bravyiEfficientAlgorithmsMaximum2014}, whereas an efficient list decoder can be done in $\calO(Ln\log n)$ \cite{Tal_2015} (the inclusion of $T$ increases the complexity, but it is still dominated by $\calO(Ln\log n)$).  Given the similar performance to $\chi = 4,6$ with $L = 8,32$, we are able to also claim an improved decoding complexity. 

Additionally, to demonstrate the improvement of the stages of the algorithm, Figure \ref{64stages} plots the $[\!]64,2,L=4\rrbracket$ code for the initial $\mathcal{A}_0$ (based on \cite{gongImprovedLogicalError2024}), the best found $\mathcal{A}$, the best found $T$ for the given $\mathcal{A}$, and the optimized pair $(T,\mathcal{A})$. The final code has similar performance to the unoptimized $\llbracket128,2,L=4\rrbracket$, despite $N$ being half as large.

\begin{figure*}[h]
    \centering
    {\input{N2L4.pgf}}
\vspace{-3mm}
    \caption{The $\llbracket 64,2,L=4\rrbracket$ code seen through the stages of \mbox{Algorithm \ref{alg:joint}}, and compared against the $\llbracket 128, 2, L=4 \rrbracket$ polar code. $\mathcal{A}_0$ is the profile from \cite{gongImprovedLogicalError2024}.} 
    \label{64stages}
\end{figure*}

\subsection{A remark on $T$}

The involution requirement of $T$ is more easily satisfied when $T$ is sparse, so the search is constrained so that off-diagonal elements $T_{i,j}$ can only be non-zero when $i$ is an information bit and $j$ is a frozen bit. From the work in \cite{kannPathMetricBased2024}, it appears that these indices of $T$ matter the most. However, the persymmetric requirement means that for every decision on $T_{i,j}$, we must make the same decision on $T_{N-j-i, N-1-i}$. For the stabilizer bits, \mbox{$i \in \mathcal{A} \iff N-1-i \in \mathcal{F}$} and \mbox{$j \in \mathcal{F} \iff N-j-i \in \mathcal{A}$}. This means that the persymmetry does not affect this imposed constraint. While the ultimate goal is to protect the logicals, works from \cite{kannPathMetricBased2024, TailBitingPAC} show that one error impacts the likelihood decisions of future bits, meaning that when the decoder better protects \emph{stabilizer} bits, the decoder will better able to decode \emph{logical} bits correctly. Naturally, it  seems beneficial to directly protect the logical bits with a precoding. However, for the logical pairs, modifying $T_{i,j}$ modifies $T_{N-j-i, N-1-i}$, but now $N-j-i$ \emph{and} $N-1-i$ are in $\mathcal{A}$. In these instances, we do have the decision of previous \emph{stabilizer} bits affect future logical bits. Thus, there seems to be some tension between the protection of one logical at the cost of degrading the protection of its pair. 

\subsection{Gates and Stabilizer weights}

While a hurdle for quantum polar codes becoming practical is their large weight stabilizers, we nevertheless discuss the impact of the precoding on the construction of quantum circuit. 

For the purposes of illustration, imagine we have $T = \begin{bmatrix}
    1 & 1\\ 0 & 1
\end{bmatrix}$, a reversed CNOT gate, and $G = \begin{bmatrix}
    1 & 0 \\ 1 & 1
\end{bmatrix}$, a CNOT gate. Combined, $TG = \begin{bmatrix} 0 & 1 \\ 1 & 1\\ \end{bmatrix}$, which can be implemented exactly as a reversed CNOT followed by a CNOT, seen as  
\begin{align*}
    (a,b) &\rightarrow (a, a\oplus b) \\
    (a, a\oplus b) &\rightarrow (a\oplus a \oplus b, a \oplus b) = (b, a \oplus b), 
\end{align*}
but cannot be implemented as just one gate. Therefore an off-diagonal element of $T$ directly translates to an additional Clifford gate. However, for the syndrome measurement, we are only concerned with the non-zero elements in the stabilizer matrix. This means an off-diagonal element of $T$ modifies $F^TT^TG^T$ in a way that may not directly translate to an additional Clifford gate. For example, looking at the $\llbracket64,2\rrbracket$ construction in Figure \ref{64stages}, $||T||_0 = 100$, meaning there are $36$ off-diagonal elements, requiring $36$ additional gates on the encoders side. However, $|| F^TT^TG^T||_0 - ||F^TG^T||_0 = 26$. This means that the number of gates required going from quantum polar codes to quantum precoded polar codes is relatively small. In terms of total gates required, looking at the $\llbracket 256, 2 \rrbracket$ code, $||F^TT^TG^T||_0 = 5550$, meaning the syndrome extraction would require $11100$ Clifford gates (ignoring the basis-change gates). This is around a $2.3$x increase over the $4704$ Clifford gates for the surface code. 

\section{Conclusion}
\label{conclusion}
We have presented quantum precoded polar codes, a new family of codes that apply the breath of literature on short blocklength polar codes towards quantum polar codes. Analysis shows how a precoder can be implemented for encoding and syndrome extraction, as well as the algorithmic changes the decoder makes. A search finds codes of length $N = 256$ and $512$ that compare favorably against the surface code in terms of logical error rate, with fewer encoding/ancilla qubits required, larger rate, and better complexity, but at the cost of an increased number of Clifford gates required. These codes may be useful in a regime where increasing Clifford gates is easier than increasing the number of qubits. 

\section*{Acknowledgment}

We are grateful to Rüdiger Urbanke for helpful discussions.

\newpage


\IEEEtriggeratref{14}
\bibliographystyle{IEEEtranNoURL}
\bibliography{references.bib}



\end{document}